%% file: Robust_Partial_Least_Squares_via_Low_Rank_and_Sparse_Decomposition.tex
\documentclass[10pt,journal]{IEEEtran}
\usepackage{times}
\usepackage{amssymb}
\usepackage{enumitem}
\usepackage[utf8]{inputenc} 
\usepackage[T1]{fontenc}    
\usepackage{hyperref}       
\usepackage{url}            
\usepackage{booktabs}       
\usepackage{amsfonts}       
\usepackage{nicefrac}       
\usepackage{microtype}      
\usepackage{color}
\usepackage{epsfig}
\usepackage{graphicx}
\usepackage{amsmath}
\usepackage{amssymb}
\usepackage{subfigure}
\usepackage{multirow}
\usepackage{array}
\usepackage{algorithm}
\usepackage{algorithmicx}
\usepackage[utf8]{inputenc}
\usepackage{algpseudocode}
\usepackage{amsbsy}
\usepackage{esint}
\usepackage{float}
\usepackage{graphics}

\graphicspath{{./figures/}}

\input{Ahmed-defs}

\def\t{{\vec t}}
\def\w{{\vec w}}
\def\T{{\mtx Q}}
\def\P{{\mtx \Lambda_x}}
\def\C{{\mtx \Lambda_y}}
\def\E{{\mtx \Delta_x}}
\def\F{{\mtx \Delta_y}}
\def\L{{\mtx L}}
\def\M{{\mtx M}}
\def\I{{\mtx I}}
\def\X{{\mtx X}}
\def\Y{{\mtx Y}}
\def\A{{\mtx A}}
\def\B{{\mtx B}}
\def\D{{\mtx D}}
\def\U{{\mtx U}}
\def\V{{\mtx V}}
\def\S{{\mtx S}}

\def\J{{\mtx J}}
\def\W{{\mtx W}}

\makeatother
\title{Robust Partial Least Squares Using Low Rank and Sparse Decomposition}
\author{
\begin{tabular}{*{2}{>{\centering}p{.5\textwidth}}}
\large Farwa Abbas & \large Hussain Ahmad \tabularnewline
\textit{Department of Electrical and Electronic Engineering} & \textit{School of Computer and Mathematical Sciences} \tabularnewline
\textit{Imperial College London} & \textit{University of Adelaide, Australia} \tabularnewline
f.abbas20@imperial.ac.uk & hussain.ahmad@adelaide.edu.au 
\end{tabular}
}

\begin{document}
\maketitle

\maketitle
\begin{abstract}
This paper proposes a framework for simultaneous dimensionality reduction and regression in the presence of outliers in data by applying low-rank and sparse matrix decomposition. For multivariate data corrupted with outliers, it is generally hard to estimate the true low dimensional manifold from corrupted data. The objective of the proposed framework is to find a robust estimate of the low dimensional space of data to reliably perform regression. The effectiveness of the proposed algorithm is demonstrated experimentally for simultaneous regression and dimensionality reduction in the presence of outliers in data.
\end{abstract}
   
\section{Introduction}

In classical Multivariate Linear Regression (MLR), the objective is to model relationship between response variables $\Y$ and predictors $\X$ by fitting a linear map $\mathbf{\Theta}$ to observed samples such that $\Y-\F=\X\mathbf{\Theta}$ where $\F$ is the measurement error. The solution to this problem i.e. $\hat{\mathbf{\Theta}}=(\X^\Trsp \X)^{-1}\X^\Trsp\Y$ can be found by solving the following program.
\begin{align}
\min_{\mtx \F, \mtx \Theta}  \Vert \F \Vert_F^2 \\
\notag
\text{ s.t. } \Y-\F=\X\mathbf{\Theta}.
\end{align}
However, if the model is also prone to error then accounting for both the measurement and modelling errors in the equation $\Y-\F=(\X-\E)\mathbf{\Theta}$ using Total Least Squares (TLS) corresponds to solving the objective mentioned below.
\begin{align}
\min_{\F, \E, \mtx \Theta}  \Vert \F \Vert_F^2 + \Vert \E \Vert_F^2 \\
\notag
\text{ s.t. } \Y-\F=(\X-\E)\mathbf{\Theta}.
\end{align}
The above program tries to find a linear map from $\X$ to $\Y$ to model the relationship between predictors and responses while penalizing the errors in both $\X$ and $\Y$. In most practical cases, there are strong correlations between subsets of predictors. Therefore, they tend to span a low dimensional subspace and hence the covariance matrix turns out to be singular, making the linear regression problem ill-conditioned \cite{abc}. To resolve this problem, one possible solution is to project each sample onto the low-dimensional subspace spanned by uncorrelated predictors and form a smaller set of new orthogonal predictors, each being the linear combination of original set of predictors. This can be done by finding singular value decomposition of $\X$ such that $\X=\T\P^\Trsp$ where $\T$ and $\P$ are the scores and loadings matrices respectively. The new set of orthogonal predictors in $\T$ can be regressed to $\Y$ by regression equation $\Y=\T\mathbf{\Theta}+\F$ and the solution to this new problem is $\tilde{\mathbf{\Theta}}=(\T^\Trsp \T)^{-1}\T^\Trsp\Y$. It must be noted that our new covariance matrix is now invertible even in the presence of strong correlations between predictors. This is the main idea behind Principal Component Regression (PCR) \cite{park1981collinearity}. However PCA, being an unsupervised dimensionality reduction approach, computes latent components while completely disregarding the information obtained from responses and hence, the predictive information can probably be discarded and noise may be preserved inducing a poor regression model \cite{geladi1986partial}.

In order to estimate the low dimensional subspace by taking into account the information provided by both predictors and responses, we aim to find a low rank approximation of both $\X$ and $\Y$ by solving the following optimization problem.
\begin{align}
\label{eqq3} 
\min_{\F, \E, \mtx \Theta}  \Vert \F \Vert_F^2 + \Vert \E \Vert_F^2 + \Vert \T\P^\Trsp \Vert_* + \Vert \T\C^\Trsp \Vert_* \\
\notag
\text{ s.t. } \X-\E=\T\P^\Trsp, \Y-\F=\T\C^\Trsp. \hspace{1cm} 
\end{align}
We can argue that Partial Least Squares (PLS) \cite{tenenhaus1998regression} is an iterative algorithm that fundamentally tries to solve the above mentioned objective by maximizing the covariance between predictors and responses. It aims to find the first latent component $\t =\X \w$ by finding weights $\w$ while maximizing the covariance  $\cov(\X \w,\Y)$ subject to $\w^\Trsp \w=1$. The subsequent latent components are found by singular value decomposition of residual matrix that is updated iteratively. Unlike conventional PLS method that computes the scores matrix as $\T = \X\W(\P^\Trsp \W)^{-1}$, we observe that it can rather be estimated in a flexible way by optimizing the objective in (\ref{eqq3}) without imposing any explicit form. Unlike PCA that induces an orthogonal projection $\T(\T^\Trsp \T)^{-1}\T^\Trsp$ matrix, PLS algorithm instead induces an oblique projection $\W(\P^\Trsp \W)^{-1}\P^\Trsp$ matrix in a supervised way \cite{phatak1997geometry}.  We argue that the optimization program in (\ref{eqq3}) tries to find the low dimensional manifold similar to the one found by PLS iterative algorithm without finding any oblique projections. Moreover, we claim that we can achieve the same objective without requiring a weight matrix $\W$ for regression or dimensionality reduction.

\section{Related Work}
PLS algorithm has widely been used in chemometrics \cite{mani2019investigating, sampaio2018optimization}, clinical chemistry \cite{hall1992near, grafen2016performance}, industrial process control \cite{chen2016pca, dong2015adaptive, li2016big} and many other applications \cite{casal1996comparative,sanchez2006exploring, einax1998assessing}. However, its effectiveness is restricted due its extreme sensitivity to outliers in data \cite{gil1998robust}. Several methods have been proposed in past decades to develop a robust PLS model that include \cite{gil1998robust, hubert2003robust, gonzalez2009robust, wakelinc1992robust, hoffmann2016sparse}.  Primarily, these methods have proposed to use a robust estimate of covariance matrix to mitigate the effect of outliers in data. Two most commonly used methods to estimate a robust covariance matrix are Orthogonalized Gnanadesikan-Kettenring (OGK) \cite{maronna2002robust} and Minimum Covariance Determinant (MCD) \cite{rousseeuw1999fast}. However, the utility of these methods is restricted to the cases where the number of samples are very large as compared to the number of predictors which is unrealistic in many situations \cite{boudt2017minimum}. 

It can be observed that, instead of minimizing $\Vert \E \Vert_F^2$ and $\Vert \F \Vert_F^2$ as in conventional PLS method, we can minimize instead $\Vert \E \Vert_1$ and $\Vert \F \Vert_1$ to get rid of outliers in data.  Motivated by this observation, we propose to employ a new low-rank and sparse decomposition framework for simultaneously decomposing both predictor and response matrices. We refer to our proposed algorithm as Robust PLS (RPLS). We have shown experimentally that by allowing sparse errors in both predictors and responses, the low-dimensional space can be discovered reliably in the presence of outliers.

 
In contrast to iterative PLS scheme, we propose a new way to find the decompositions $\X = \T\P^\Trsp + \E$ and $\Y = \T\C^\Trsp + \F$ that does not involve any power method or oblique projections at all. We introduce a novel framework to perform regression by finding projections onto the latent space estimated by a robust method. More precisely, we make the assumption that after removing errors in both $\X$ and $\Y$, the low rank matrices $\X-\E = \T\P^\Trsp$ and $\Y-\F = \T\C^\Trsp$ can be related through regression equation as $\T\C^\Trsp = \T\P^\Trsp \mathbf{\Theta}$. Moreover, if $\T$ is constrained to have orthonormal columns i.e. $\T^\Trsp \T = \I$ then our regression equation reduces to $\C^\Trsp = \P^\Trsp\mathbf{\Theta}$. Our method is straightforward and also handles outliers in data. We show the usefulness of proposed framework for simultaneous regression and dimensionality reduction. We, further, show that by performing regression after projecting onto the low dimensional space, the response variable can be predicted in a reliable and robust way. 

\subsection{Our Contributions}
\label{ssec: contrib}
In this paper, we propose a novel method to formulate PLS regression problem by introducing a low-rank and sparse decomposition framework. To the best of our knowledge, this is the first work that solves the PLS regression problem in an optimization framework by effectively penalizing the outliers in data. In particular, we make the following contributions:
\begin{enumerate}
\item We propose a novel way to formulate PLS regression problem as a low rank and sparse decomposition problem by minimizing the proposed objective using Alternating Direction Method of Multipliers (ADMM).
\item We propose a regression approach based on projection onto a latent space.
\item Finally, we demonstrate experimentally the superior performance of our proposed method on both synthetic and real world datasets corrupted with outliers.
\end{enumerate}


\section{Mathematical Formulation}
\label{sec:format}

Motivated by the approach discussed in \cite{shang2014recovering}, we develop a robust PLS model in low-rank and sparse decomposition framework. More formally, we aim to find $\T \in \mathbb{R}^{n \times k}$ (such that $\T^\Trsp \T = \I$), $\P \in \mathbb{R}^{p \times k}$ and $\C \in \mathbb{R}^{r \times k}$ such that $\T\P^\Trsp \in \mathbb{R}^{n \times p}$ and $\T\C^\Trsp \in \mathbb{R}^{n \times r}$ are low-rank matrices where $k$ is an upper bound on the rank of $\X$ and $\Y$ i.e., $\text{rank}(\X) \leq k$ and $\text{rank}(\Y) \leq k$. We aim to find a low rank approximation for $\X$ and $\Y$ while ensuring the error matrices $\E$ and $\F$ to be sparse. The problem can be mathematically formulated as follows.
\begin{small}\begin{align}
\label{eq:obj}
\notag 
&\min_{\T,\P,\C,\E,\F}\Vert \E \Vert_1 +\Vert \F \Vert_1 +\lambda_1 \Vert \T\P^\Trsp \Vert_* + \lambda_2 \Vert \T\C^\Trsp \Vert_* \\ 
&\text{ subject to }\X=\T\P^\Trsp +\E \text{ , } \Y=\T\C^\Trsp +\F \text{ and } \T^\Trsp \T=\I.
\end{align}\end{small}
In this paper, we restrict our attention to solve the optimization program in (\ref{eq:obj}). The partial augmented Lagrangian function of the program in (\ref{eq:obj}) can be written as follows.
\begin{small}\begin{align}
\notag
&\mathcal{L}(\T,\P,\C,\E,\F,\L,\M) =\Vert \E \Vert_1 + \lambda_1 \Vert \P \Vert_*+\Vert \F \Vert_1,  \\ \notag
&+ \lambda_2 \Vert \C \Vert_* + \langle \L,\X-\T\P^\Trsp -\E \rangle +\frac{\alpha_1}{2}\Vert \X-\T\P^\Trsp -\E \Vert_F^2,\\
&+ \langle \M,\Y-\T\C^\Trsp -\F \rangle+\frac{\alpha_2}{2}\Vert \Y-\T\C^\Trsp -\F \Vert_F^2,
\end{align}\end{small}
where $\L$ and $\M$ are the matrices of Lagrange multipliers and $\alpha_1$ and $\alpha_2$ are penalty parameters for first two constraints. The last constraint regarding the orthogonality of $\T$ is enforced separately while computing the update for $\T$ . $\lambda_1$ and $\lambda_2$ are weights of the second and fourth terms respectively.

The objective function defined above is non-convex therefore we resort to alternative minimization through non-convex ADMM. By keeping all other matrices to be fixed, the update for each individual matrix in the objective function can be obtained. First, the update for $\T$ can be computed while keeping all other matrices fixed.
\begin{small}\begin{align*}
\T&^{(k+1)}=\arg \min_\T \mathcal{L}(\T,\P^{(k)},\C^{(k)},\E^{(k)},\F^{(k)},\L^{(k)},\M^{(k)}) \\
&=\arg \min_\T \langle \L,\X-\T\P^\Trsp -\E \rangle+\frac{\alpha_1}{2}\Vert \X-\T\P^\Trsp -\E \Vert_F^2 \\ 
&+\langle \M,\Y-\T\C^\Trsp -\F \rangle+\frac{\alpha_2}{2}\Vert \Y-\T\C^\Trsp -\F \Vert_F^2 \\
&=\arg \min_\T \frac{\alpha_1}{2}\Big \Vert\frac{\L}{\alpha_1}+(\X-\T\P^\Trsp -\E)\Big \Vert_F^2 \\
&+ \frac{\alpha_2}{2}\Big \Vert\frac{\M}{\alpha_2}+(\Y-\T\C^\Trsp -\F)\Big \Vert_F^2 \\
&=\arg \min_\T \frac{\alpha_1}{2}\Big \Vert(\frac{\L}{\alpha_1}+\X-\E)-\T\P^\Trsp \Big \Vert_F^2 \\
&+\frac{\alpha_2}{2}\Big \Vert(\frac{\M}{\alpha_2}+\Y-\F)-\T\C^\Trsp \Big \Vert_F^2 \\
&=\arg \min_\T \frac{\alpha_1}{2}\Vert \B-\T\P^\Trsp \Vert_F^2+\frac{\alpha_2}{2}\Vert \A-\T\C^\Trsp \Vert_F^2.
\end{align*}\end{small}
By substituting $\B=\frac{\L}{\alpha_1}+\X-\E$ and $\A=\frac{\M}{\alpha_2}+\Y-\F$ above we get the last equation. By expanding the squared terms we get the following.
\begin{align*}
\T^{(k+1)}&= \arg \min_\T -\alpha_1 \langle \B, \T\P^\Trsp \rangle -\alpha_2\langle \A, \T\C^\Trsp  \rangle \\
&= \arg \max_\T \alpha_1 \langle \B, \T\P^\Trsp \rangle +\alpha_2\langle \A, \T\C^\Trsp  \rangle \\
&= \arg \max_\T \langle\alpha_1\B\P, \T\rangle +\langle \alpha_2\A\C, \T \rangle \\
&= \arg \max_\T \langle\alpha_1\B\P+ \alpha_2\A\C, \T \rangle \\
&= \arg \max_\T \langle \D, \T \rangle,
\end{align*}
where $\D= \alpha_1\B\P+ \alpha_2\A\C$. Now we want to minimize the following objective while simultaneously enforcing $\T^\Trsp \T=\I$.
\begin{align*}
\T^{(k+1)}&= \arg \max_\T \langle \D, \T \rangle  \\
&= \arg \max_\T \langle \U\S\V^\Trsp , \T \rangle \\
&= \arg \max_\T \langle \S, \U^\Trsp \T\V \rangle \\
&= \arg \max_\T \langle \S, \J \rangle \hspace{1.3cm} \text{ s.t. } \T^\Trsp \T=\I.
\end{align*}
In the set of equations above $\D=\U\S\V^\Trsp$ is the singular value decomposition of matrix $\D$. The matrix $\J=\U^\Trsp \T\V$ is an orthonormal matrix, as it is a product of orthonormal matrices, and thus the expression is maximised when $\J$ equals identity matrix $\I$. Thus
\begin{align}
\notag 
\J&=\U^\Trsp \T\V=\I \\
\T^{(k+1)}&=\U\V^\Trsp.
\label{eq:2}
\end{align}
Similarly, the updates for $\P$ and $\C$ can be computed as follows.
\begin{align}
\notag 
\P^{(k+1)},\C^{(k+1)}&=\arg \min_{\P,\C} \frac{\lambda_1}{\alpha_1}\Vert \P \Vert_* +\frac{1}{2}\Vert \B^\Trsp \T-\P \Vert_F^2 \\ \notag
& + \frac{\lambda_2}{\alpha_2}\Vert \C \Vert_* +\frac{1}{2}\Vert \A^\Trsp \T-\C \Vert_F^2 \\
\P^{(k+1)}&=\tilde{\U}\mathcal{S}_{\lambda_1/\alpha_1}(\tilde{\mathbf{\Sigma}}) \tilde{\V}^\Trsp  \label{eq:3} \\
\C^{(k+1)}&=\widehat{\U}\mathcal{S}_{\lambda_2/\alpha_2}(\widehat{\mathbf{\Sigma}}) \widehat{\V}^\Trsp,
\label{eq:4}
\end{align}
where $\B^\Trsp \T = \tilde{\U}\tilde{\mathbf{\Sigma}}\tilde{\V^\Trsp }$ and $\A^\Trsp \T=\widehat{\U}\widehat{\mathbf{\Sigma}}\widehat{\V}^\Trsp$ are the singular value decompositions of $\B^\Trsp\T$ and $\A^\Trsp\T$ respectively. The above result is obtained from the proximal operator of nuclear norm minimization problem where $\mathcal{S}_\epsilon$ is the soft thresholding operator as defined in \eqref{eq:e}.
\begin{align}
\mathcal{S}_\epsilon(\mtx K)&=\begin{cases}K_{ij}-\epsilon & K_{ij}>\epsilon\\K_{ij}+\epsilon & K_{ij}<\epsilon\\0& \text{otherwise}\end{cases} \label{eq:e}
\end{align}
By keeping all the other matrices to be fixed we can find the update equations for $\E$ and $\F$ by using the proximal operator of $\ell_1$ norm minimization problem.
\begin{small}
\begin{align}
\notag
&\E^{(k+1)},\F^{(k+1)}=\arg \min_{\E,\F}  +\frac{1}{2}\Big \Vert(\X-\T\P^\Trsp +\frac{\L}{\alpha_1} )-\E \Big \Vert_F^2 \\ \notag
& +\frac{1}{2} \Big \Vert(\Y-\T\C^\Trsp +\frac{\M}{\alpha_2} )-\F \Big \Vert_F^2+\frac{1}{\alpha_1} \Vert \E \Vert_1 + \frac{1}{\alpha_2}\Vert \F \Vert_1
\end{align}
\end{small}
\begin{align}
\E^{(k+1)}&=\mathcal{S}_{1/\alpha_1} \left(\X-\T^{(k+1)}\P^{(k+1)\Trsp } +\frac{\L^{(k)}}{\alpha_1} \right) \label{eq:5} \\
\F^{(k+1)}&=\mathcal{S}_{1/\alpha_2} \left(\Y-\T^{(k+1)}\C^{(k+1)\Trsp } +\frac{\M^{(k)}}{\alpha_2} \right) \label{eq:6}
\end{align}
In the end the matrices of Lagrange multipliers are updated as follows.
\begin{align}
\L^{(k+1)} &= \L^{(k)}+\alpha_1^{(k)}(\X-\T^{(k+1)}\P^{(k+1)\Trsp } -\E^{(k+1)}) \label{eq:7}\\
\M^{(k+1)} &= \M^{(k)}+\alpha_2^{(k)}(\Y-\T^{(k+1)}\C^{(k+1)\Trsp } -\F^{(k+1)}). \label{eq:8}
\end{align}
The complete algorithm is shown in Algorithm 1. Here $\hat{\X}^{(k+1)}$ and $\hat{\Y}^{(k+1)}$ are the approximations of $\X$ and $\Y$ at iteration $k+1$ i.e. $\hat{\X}^{(k+1)} = \T^{(k+1)}\P^{(k+1)\Trsp } +\E^{(k+1)}$ and $\hat{\Y}^{(k+1)} = \T^{(k+1)}\C^{(k+1)\Trsp } +\F^{(k+1)}$.
\begin{algorithm} [t]
\caption{RPLS via ADMM}
\textbf{Input:} $\lambda_1, \lambda_2, tol$ \\
\textbf{Initialization:} $\alpha_1^{(0)}, \alpha_2^{(0)}, \alpha_{max}, \rho, \T^{(0)}=\operatorname{eye}(n,k),$ \\ $\P^{(0)}=\mtx 0, \C^{(0)}=\mtx 0, \E^{(0)}=\mtx 0, \F^{(0)}= \mtx 0$ \\
\textbf{Output:} $\T$,$\P$,$\C$,$\E$,$\F$ 
\begin{algorithmic}
\While {not converged}
\State Update $\T^{(k+1)}$ by (\ref{eq:2})
\State Update $\P^{(k+1)}$ by (\ref{eq:3})
\State Update $\C^{(k+1)}$ by (\ref{eq:4})
\State Update $\E^{(k+1)}$ by (\ref{eq:5})
\State Update $\F^{(k+1)}$ by (\ref{eq:6})
\State Update $\L^{(k+1)}$ by (\ref{eq:7})
\State Update $\M^{(k+1)}$ by (\ref{eq:8})
\State Update $\alpha_1^{(k+1)}$ by \\ \hspace{1 cm} $\alpha_1^{(k+1)}=\min(\rho \alpha_1^{(k)},\alpha_{max})$
\State Update $\alpha_2^{(k+1)}$ by \\ \hspace{1 cm} $\alpha_2^{(k+1)}=\min(\rho \alpha_2^{(k)},\alpha_{max})$
\State Check convergence by \\
$\Vert \X-\hat{\X}^{(k+1)} \Vert_F +\Vert \Y-\hat{\Y}^{(k+1)} \Vert_F <$  $tol$
\EndWhile
\end{algorithmic}
\end{algorithm}
\section{COMPUTATION COMPLEXITY ANALYSIS}
\label{sec:pagestyle}

In the above algorithm the computational cost in all the updates is primarily due to svd and matrix multiplications. 
In equation \eqref{eq:2}, for matrix $\D \in \mathbb{R}^{n \times k}$ the computational complexity of full svd is $\mathcal{O}(nk^2)$ $(n > k)$. Similarly the next two updates \eqref{eq:3} , \eqref{eq:4} involve svd having costs  
$\mathcal{O}(pk^2)$, and $\mathcal{O}(rk^2)$ respectively. The cost of matrix multiplications $\T^{(k+1)}\P^{(k+1)\Trsp}$ in \eqref{eq:5} and \eqref{eq:7} is $\mathcal{O}(nkp)$. Similarly, the cost associated with $\T^{(k+1)}\C^{(k+1)\Trsp}$ in \eqref{eq:6} and \eqref{eq:8} is $\mathcal{O}(nkr)$. Hence the total cost is $\mathcal{O}(T*((k^2(n+p+r)+2nk(p+r))))$ where $T$ is total number of iterations. 

\section{EXPERIMENTAL DETAILS}
\label{sec:typestyle}
In this section we compare the performance of our proposed algorithm on both synthetic and real world datasets. The experiments on synthetic data show the performance on dimensionality reduction and the real dataset is evaluated for both dimensionality reduction and regression. All the experiments have been performed on MATLAB 2018a with Core i5 CPU (2.50 GHz), 8 GB RAM and Windows 10 operating system. 

\subsection{Results on Synthetic dataset}
For randomly generated synthetic data, the predictor matrix is generated randomly having dimensions $150 \times 40$ i.e. we have $150$ samples and $40$ predictors. The number of orthogonal components is chosen to be $5$. To simulate multicollinearity in data, out of $40$ predictors, some of the predictors are linearly combined to make other predictors. Sample size is deliberately made larger as compared to the number of predictors so that existing methods for robust PLS can be applied. We compare our results with traditional PLS and PLS models robustified by using frequently used robust estimates of covariance matrix i.e. PLS-OGK and PLS-MCD. Four responses are being generated from linear functions of randomly selected predictors and are then stacked in a $150 \times 4$ dimensional response matrix. 

\begin{figure}[t]
\begin{minipage}[b]{1.0\linewidth}
  \centering
  \centerline{\includegraphics[width=8.5cm]{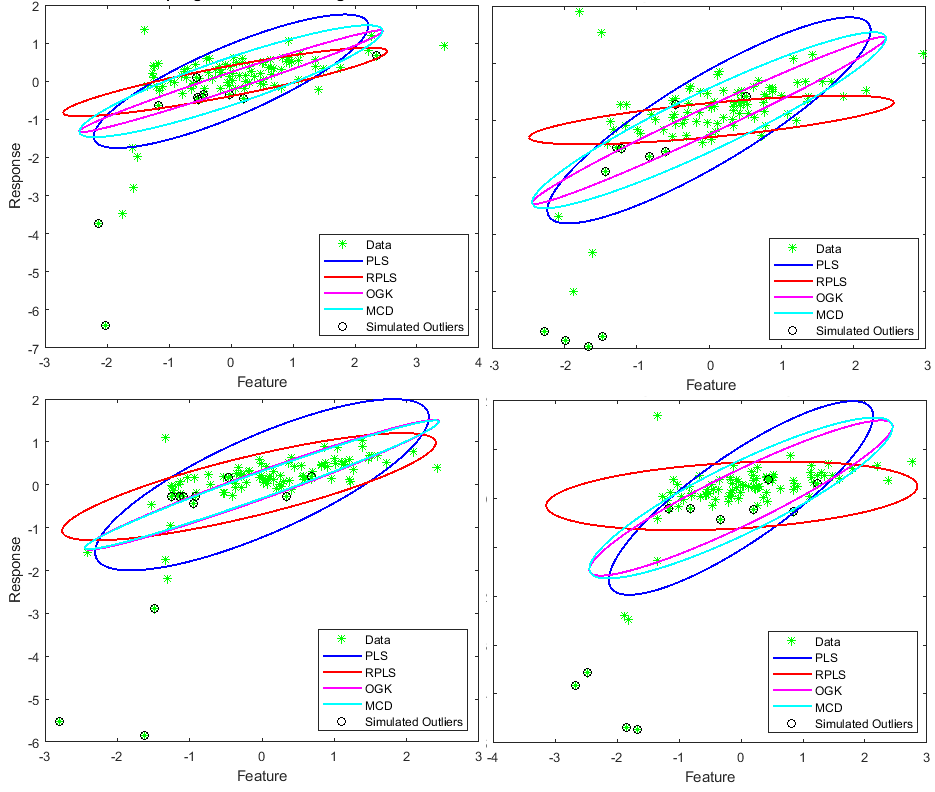}}
\caption{Results of PLS, PLS-OGK, PLS-MCD and RPLS on four times randomly generated synthetic data that is corrupted by outliers. RPLS is able to correctly determine the low-dimensional space in all cases.}
\label{synth}
\end{minipage}
\end{figure}

To simulate the effect of presence of outliers, we have introduced two different types of outliers in data. First ones are the sparse outliers, that are added randomly to some of the predictors and responses. Second type of outliers perturb lowest $10\%$ values in the response to $10$ times their value. To increase reliability, this experiment has been repeated four times and results for all four experiments are shown in Figure \ref{synth}. By introducing outliers, the traditional PLS algorithm fails to correctly discover the low dimensional structure in data as depicted in Figure \ref{synth}. The plotted ellipses show the region of $95\%$ confidence determined by different algorithms.
\subsection{Results on Real dataset}
Partial least squares regression has extensively been applied in spectroscopy to reveal the spectral content of various chemical compositions. In past decades, PLS has achieved most promising results to predict the octane number of gasoline using near infrared (NIR) spectroscopic techniques \cite{mendes2012determination, ozdemir2005determination}.
We test the performance of our algorithm on a dataset containing the NIR spectra of 60 gasoline samples with known octane numbers and covers the range from 900 to 1700 nm in 2 nm intervals. The size of complete predictor matrix is $60 \times 401$ and number of orthogonal components are chosen to be $10$. 

\begin{figure}[t]

\begin{minipage}[b]{1.0\linewidth}
  \centering
  \centerline{\includegraphics[width=6cm]{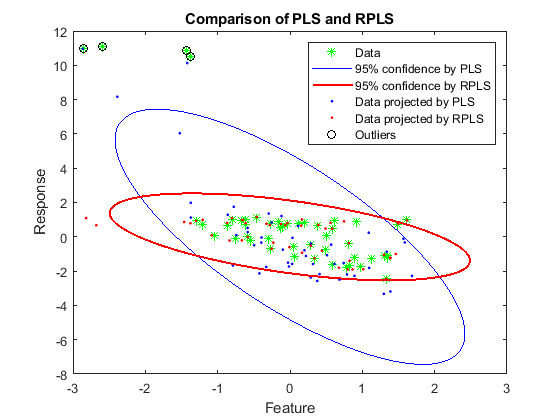}}
\caption{RPLS has correctly identified the underlying manifold representing the spectral measurements of gasoline samples.}
\label{spectra}
\end{minipage}
\end{figure}

Since the number of samples are very less as compared to the dimensionality of data, therefore, PLS-OGK and PLS-MCD could not be applied. The data has been corrupted with second type of outliers i.e. outliers introduced by adding a perturbation on $10\%$ lowest values and results produced by both PLS and RPLS are compared in Figure \ref{spectra}.

After learning the PLS model with $80\%$ of the data, the remaining $20\%$ data is tested to predict the response. Results for regression are compared for MLR, PCR, and PLSR by using their original regression equations. The proposed regression results i.e. regression by projection are also compared for both PLS and RPLS. All these results are summarized in Table \ref{tab:Tab1}. It can be seen that MLR is unable to produce reliable estimate of the response. PCR produces better results but it completely disregards information in response while finding projections. PLS projects the data in a supervised way but its estimates are degraded due to outliers. RPLS performs well as compared to all three since it finds robust projections in a supervised manner. NMSE defined by $\Vert \Y_{true}-\Y_{est} \Vert_2/\Vert \Y_{true}\Vert_2$ can be compared for predictions obtained below. 

\begin{table}[H]
\centering
\resizebox{1\columnwidth}{!}{%
\begin{tabular}{cccccc}
\hline
\textbf{Octane Num} & \textbf{MLR}      & \textbf{PCR}      & \textbf{PLS}      & \textbf{PLSR}     & \textbf{RPLS}   \\ \hline
88.45                                & 74.83                              & 87.97                              & 81.61                              & 94.73                              & 90.12                            \\ \hline
88.70                                & 111.46                             & 89.23                              & 88.48                              & 95.41                              & 89.09                            \\ \hline
88.10                                & 32.23                              & 91.80                              & 110.69                             & 89.62                              & 88.22                            \\ \hline
87.60                                & 80.89                              & 86.17                              & 85.28                              & 83.17                              & 87.20                            \\ \hline
88.35                                & 65.94                              & 90.63                              & 87.10                              & 89.13                              & 88.43                            \\ \hline
85.10                                & 112.51                             & 87.16                              & 91.89                              & 86.98                              & 85.05                            \\ \hline
85.10                                & 108.17                             & 87.78                              & 76.13                              & 87.75                              & 85.27                            \\ \hline
84.70                                & 69.04                              & 84.02                              & 72.18                              & 82.19                              & 83.92                            \\ \hline
87.20                                & 103.74                             & 94.09                              & 81.34                              & 91.61                              & 86.67                            \\ \hline
86.60                                & 101.65                             & 87.13                              & 85.44                              & 81.50                              & 85.57                            \\ \hline
89.60                                & 106.25                             & 88.93                              & 91.66                              & 81.61                              & 88.84                            \\ \hline
87.10                                & 79.30                              & 86.46                              & 94.23                              & 82.25                              & 87.66                            \\ \hline
NMSE       & 0.271670 & 0.029946 & 0.101360 & 0.053032 & \textbf{0.0081} \\ \hline
\end{tabular}
}
\caption{\label{tab:Tab1}Predictions obtained by MLR, PCR, PLS, PLSR, and RPLS.}
\end{table}

%

\section{Conclusion}
\label{sec:majhead}
This paper proposed a novel framework for dimensionality reduction and regression by finding a low-rank and sparse decompositions for both predictors and responses. Experiments have shown that the proposed algorithm performs better as compared to the other techniques in presence of outliers and corruptions. We further show that our method approximates the regression function better than the existing algorithms and hence achieves smallest error while making predictions on unseen data in the presence of outliers.

\vfill\pagebreak
{\small
\bibliographystyle{unsrt}
\bibliography{refs}
}

\end{document}

%% file: Ahmed-defs.tex
\newcommand{\C}{\mathbb{C}}

\newcommand{\E}{\operatorname{E}}
\newcommand{\F}{\mathrm{F}}
\newcommand{\cov}{\operatorname{cov}}

\newcommand{\mtx}[1]{\boldsymbol{#1}}

\newcommand{\Trsp}{\mathrm{T}}




%% file: Robust_Partial_Least_Squares_via_Low_Rank_and_Sparse_Decomposition.bbl
\begin{thebibliography}{10}

\bibitem{abc}
Herman Wold.
\newblock Nonlinear estimation by iterative least squares procedures.
\newblock {\em F. N. David and J. Neyman, Eds., Research Papers in Statistics, Festschrift for J. Neyman}, 1966.

\bibitem{park1981collinearity}
Sung~H Park.
\newblock Collinearity and optimal restrictions on regression parameters for estimating responses.
\newblock {\em Technometrics}, 23(3):289--295, 1981.

\bibitem{geladi1986partial}
Paul Geladi and Bruce~R Kowalski.
\newblock Partial least-squares regression: a tutorial.
\newblock {\em Analytica chimica acta}, 185:1--17, 1986.

\bibitem{tenenhaus1998regression}
Michel Tenenhaus.
\newblock {\em La r{\'e}gression PLS: th{\'e}orie et pratique}.
\newblock Editions technip, 1998.

\bibitem{phatak1997geometry}
Aloke Phatak and Sijmen De~Jong.
\newblock The geometry of partial least squares.
\newblock {\em Journal of Chemometrics: A Journal of the Chemometrics Society}, 11(4):311--338, 1997.

\bibitem{mani2019investigating}
Ahmad Mani-Varnosfaderani, Samaneh Ehsani, and Yadollah Yamini.
\newblock Investigating the effects of chemical composition of motor oils on their viscosity indices using gas chromatography and chemometrics techniques.
\newblock {\em Petroleum Science and Technology}, pages 1--9, 2019.

\bibitem{sampaio2018optimization}
Pedro~Sousa Sampaio, Andreia Soares, Ana Castanho, Ana~Sofia Almeida, Jorge Oliveira, and Carla Brites.
\newblock Optimization of rice amylose determination by nir-spectroscopy using pls chemometrics algorithms.
\newblock {\em Food chemistry}, 242:196--204, 2018.

\bibitem{hall1992near}
Jeffrey~W Hall and Alan Pollard.
\newblock Near-infrared spectrophotometry: a new dimension in clinical chemistry.
\newblock {\em Clinical chemistry}, 38(9):1623--1631, 1992.

\bibitem{grafen2016performance}
M~Grafen, K~Nalpantidis, D~Ihrig, HM~Heise, and A~Ostendorf.
\newblock Performance testing of a mid-infrared spectroscopic system for clinical chemistry applications utilising an ultra-broadband tunable ec-qcl radiation source.
\newblock In {\em Optical Diagnostics and Sensing XVI: Toward Point-of-Care Diagnostics}, volume 9715, page 971517. International Society for Optics and Photonics, 2016.

\bibitem{chen2016pca}
Ao~Chen, Hongpeng Zhou, Yujian An, and Wei Sun.
\newblock Pca and pls monitoring approaches for fault detection of wastewater treatment process.
\newblock In {\em 2016 IEEE 25th International Symposium on Industrial Electronics (ISIE)}, pages 1022--1027. IEEE, 2016.

\bibitem{dong2015adaptive}
Jie Dong, Kai Zhang, Ya~Huang, Gang Li, and Kaixiang Peng.
\newblock Adaptive total pls based quality-relevant process monitoring with application to the tennessee eastman process.
\newblock {\em Neurocomputing}, 154:77--85, 2015.

\bibitem{li2016big}
Tao Li, Yihai He, and Chunling Zhu.
\newblock Big data oriented macro-quality index based on customer satisfaction index and pls-sem for manufacturing industry.
\newblock In {\em 2016 International Conference on Industrial Informatics-Computing Technology, Intelligent Technology, Industrial Information Integration (ICIICII)}, pages 181--186. IEEE, 2016.

\bibitem{casal1996comparative}
V~Casal, PJ~Martin-Alvarez, and T~Herraiz.
\newblock Comparative prediction of the retention behaviour of small peptides in several reversed-phase high-performance liquid chromatography columns by using partial least squares and multiple linear regression.
\newblock {\em Analytica Chimica Acta}, 326(1-3):77--84, 1996.

\bibitem{sanchez2006exploring}
Manuel~J Sanchez-Franco.
\newblock Exploring the influence of gender on the web usage via partial least squares.
\newblock {\em Behaviour \& Information Technology}, 25(1):19--36, 2006.

\bibitem{einax1998assessing}
JW~Einax, O~Kampe, and D~Truckenbrodt.
\newblock Assessing the deposition and remobilisation behaviour of metals between river water and river sediment using partial least squares regression.
\newblock {\em Fresenius' journal of analytical chemistry}, 361(2):149--154, 1998.

\bibitem{gil1998robust}
Juan~A Gil and Rosario Romera.
\newblock On robust partial least squares (pls) methods.
\newblock {\em Journal of Chemometrics: A Journal of the Chemometrics Society}, 12(6):365--378, 1998.

\bibitem{hubert2003robust}
Mia Hubert and K~Vanden Branden.
\newblock Robust methods for partial least squares regression.
\newblock {\em Journal of Chemometrics: A Journal of the Chemometrics Society}, 17(10):537--549, 2003.

\bibitem{gonzalez2009robust}
Javier Gonz{\'a}lez, Daniel Pe{\~n}a, and Rosario Romera.
\newblock A robust partial least squares regression method with applications.
\newblock {\em Journal of Chemometrics: A Journal of the Chemometrics Society}, 23(2):78--90, 2009.

\bibitem{wakelinc1992robust}
IN~Wakelinc and HJH Macfie.
\newblock A robust pls procedure.
\newblock {\em Journal of Chemometrics}, 6(4):189--198, 1992.

\bibitem{hoffmann2016sparse}
Irene Hoffmann, Peter Filzmoser, Sven Serneels, and Kurt Varmuza.
\newblock Sparse and robust pls for binary classification.
\newblock {\em Journal of Chemometrics}, 30(4):153--162, 2016.

\bibitem{maronna2002robust}
Ricardo~A Maronna and Ruben~H Zamar.
\newblock Robust estimates of location and dispersion for high-dimensional datasets.
\newblock {\em Technometrics}, 44(4):307--317, 2002.

\bibitem{rousseeuw1999fast}
Peter~J Rousseeuw and Katrien~Van Driessen.
\newblock A fast algorithm for the minimum covariance determinant estimator.
\newblock {\em Technometrics}, 41(3):212--223, 1999.

\bibitem{boudt2017minimum}
Kris Boudt, Peter~J Rousseeuw, Steven Vanduffel, and Tim Verdonck.
\newblock The minimum regularized covariance determinant estimator.
\newblock {\em Statistics and Computing}, pages 1--16, 2017.

\bibitem{shang2014recovering}
Fanhua Shang, Yuanyuan Liu, James Cheng, and Hong Cheng.
\newblock Recovering low-rank and sparse matrices via robust bilateral factorization.
\newblock In {\em 2014 IEEE International Conference on Data Mining}, pages 965--970. IEEE, 2014.

\bibitem{mendes2012determination}
Gisele Mendes, Helga~G Aleme, and Paulo~JS Barbeira.
\newblock Determination of octane numbers in gasoline by distillation curves and partial least squares regression.
\newblock {\em Fuel}, 97:131--136, 2012.

\bibitem{ozdemir2005determination}
Durmu{\c{s}} {\"O}zdemir.
\newblock Determination of octane number of gasoline using near infrared spectroscopy and genetic multivariate calibration methods.
\newblock {\em Petroleum science and technology}, 23(9-10):1139--1152, 2005.

\end{thebibliography}
